\documentclass[%
reprint,
showpacs,
amsmath,amssymb,
aps,
]{revtex4-1}
\usepackage{amsmath}
\usepackage{cases}
\usepackage{amssymb}
\usepackage{graphicx}
\usepackage{dcolumn}
\usepackage{bm}
\usepackage{color}

\usepackage[mathlines]{lineno}

\begin{document}

\preprint{APS/123-QED}

\title{Improved Thermometer from Intermediate Mass Fragments in Heavy-Ion Collisions with Isobaric Yield Ratio Difference}

\author{Chun-Wang Ma
$^{1}$}
\thanks{Corresponding author. Email: machunwang@126.com}
\author{Tian-Tian Ding
$^{1}$}
\author{Chun-Yuan Qiao
$^{1}$}
\author{Xi-Guang Cao
$^{2}$}

\affiliation{%
$^{1}$ Institute of Particle and Nuclear Physics, Henan Normal University, Xinxiang 453007, China\\
$^{2}$ Shanghai Institute of Applied Physics, Chinese Academy of Sciences, Shanghai 201800, China\\
}

\date{\today}

\begin{abstract}
\begin{description}
\item[Background] Temperature is an important parameter in studying many important questions in heavy-ion collisions. A thermometer based on the isobaric yield ratio (IYR) has been proposed [Ma \textit{et al.}, Phys. Rev. C \textbf{86}, 054611 (2012) and Ma \textit{et al.}, \textit{ibid.}, Phys. Rev. C \textbf{88}, 014609 (2013)].
\item[Purpose] An improved thermometer ($T_{IB}$) is proposed based on the difference between IYRs. $T_{IB}$ obtained from isobars in different reactions will be compared.
\item[Methods] The yields of three isobars are employed in $T_{IB}$. The residual free energy of the three isobars are replaced by that of the binding energy. No secondary decay modification for odd $A$ fragment is used in $T_{IB}$.
\item[Results] The measured fragment yields in the 140$A$ MeV $^{40, 48}$Ca + $^{9}$Be ($^{181}$Ta) and $^{58, 64}$Ni + $^9$Be ($^{181}$Ta), the 1$A$ GeV $^{124, 136}$Xe + Pb, and the $^{112,124}$Sn + $^{112,124}$Sn reactions have been analyzed to obtain $T_{IB}$ from IMFs. $T_{IB}$ from most of the fragments in the $^{40, 48}$Ca and $^{58, 64}$Ni reactions is in the range of 0.6 MeV $ < T_{IB} < $ 3.5 MeV. $T_{IB}$ from most of the fragments in the $^{124}$Xe and $^{112,124}$Sn reactions is in the range of 0.5 MeV $ < T_{IB} < $ 2.5 MeV, while the range is 0.5 MeV $ < T_{IB} <$ 4 MeV from most of the fragments in the $^{136}$Xe reaction. In general, for most of the fragments $T_{IB}$ in the $^{40, 48}$Ca and $^{58, 64}$Ni reactions are very similar (except in the very neutron-rich fragments), and $T_{IB}$ from IMFs in the $^{124, 136}$Xe and $^{112,124}$Sn reactions is also similar. A slightly dependence of $T_{IB}$ on $A$ is found.
\item[Conclusions] Using the binding energy of the nucleus, $T_{IB}$ can be obtained without the knowledge of the free energies of fragments. In the investigated reactions, $T_{IB}$ from most of the IMFs is low.
\end{description}
\end{abstract}

\maketitle


\section{introduction}
Temperature ($T$) is one of the key questions in heavy-ion collision physics since the nuclear system experiences evolution changing from a very high temperature to a very low one to form the final fragments. At a high enough temperature, the liquid-gas phase transition can lead to nuclear disassembly \cite{YGMaZipfPRL99,YgMa99PRCTc,EOSPRC02Tc,DasPRC02Tc,YgMa04PRCTc,YgMa05PRCTc}. Using the Albergo isotopic thermometer, the temperature, its systematic dependence on the incident energy, and its evolution along the reaction time have been studied based on the yields of protons, neutrons, and some light isotopes,  \cite{YgMa05PRCTc,AlbNCA85DRT,Nato95T,Poch95PRLDR_T,JSWangT05,Wada97T,Surfling98TPRL,XiHF98Tiso,Odeh2000PRLT}. Some works have also used the isotopic thermometer to extract the temperature from fragments with larger atomic numbers, for example, carbon isotopes \cite{XiHF98Tiso,TrautTHFrg07} and intermediate mass fragments (IMFs) \cite{Odeh2000PRLT,Ma13CTP}. Besides the isotopic thermometer, many methods, such as the thermal energy \cite{Bonasera11PLBT}, excitation energy \cite{ZhouPei11T,Morr84PLBT_E,Poch85PRL_E,Nato02PRCT_E}, momentum fluctuation \cite{flucT10}, the correlation of two particle relative moment \cite{Surfling98TPRL,XiHF98Tiso}, and the kinetic energy spectra of light particles (slope temperature) \cite{Odeh2000PRLT,Westfall82PLBT_spec,Jacak83T_spec,SuJPRC12T_spec}, have also been employed and compared.

The Albergo thermometer is based on the thermodynamic model, which requires the system to be in equilibrium \cite{AlbNCA85DRT}. In the thermodynamic models, the fragment yield is mainly determined by its free energy, chemical potentials of protons and neutrons, and temperature \cite{AlbNCA85DRT,ModelFisher1,ModelFisher3,Huang10,Huang10PRCTf,Tsang07BET}. For the free energy of a nucleus at a finite temperature, a $T^2$ dependence of the coefficient has been introduced in the parametrizations formula by using density functional theory \cite{LeeBE_T10PRC}. For the temperature, it is proposed that the value of $T$ should be properly adopted for neutron-rich fragments \cite{Tsang07BET}. For IMFs, it is important to obtain the temperature in heavy-ion collisions since it is different from the temperature determined from the light particles with the Albergo thermometer \cite{XiHF98Tiso,TrautTHFrg07,Odeh2000PRLT,Ma13CTP}. Recently, an isobaric ratio methods have also been proposed to determine the temperature of IMFs. It shows that the primary IMFs reflect a temperature around 5 MeV \cite{Lin2014PRC,Liu2014PRC}, while the cold IMFs reflect much lower temperature \cite{Ma13CTP,MaCW12PRCT,MaCW13PRCT,Ma2013NST}.

Another question is that in thermodynamic models, temperature is incorporated in the probes for nuclear matter, such as nuclear symmetry energy, in heavy-ion collisions. For the many probes used to study the nuclear symmetry energy, such as the isoscaling method \cite{MBTsPRL01iso,Botv02PRCiso,Ono03PRCiso,Ono04RRCiso,SouzaPRC09isot}, the isobaric ratio method \cite{Huang10,MaCW11PRC06IYR,MaCW12EPJA,MaCW12CPL06,Ma13finite,RCIMa15}, and the IYR difference (IBD) method \cite{IBD13PRC,IBD13JPG,IBD14Ca,IBDCa48EPJA,IBD15Tgt,YMNST15,IBD15AMD,MaPLB15}, $T$ is a part of the probe and cannot be separated easily. In the IYR method, the ratio of the symmetry-energy coefficient ($a_{sym}$) to $T$ ($a_{sym}/T$) is instead used to study the symmetry coefficient of the neutron-rich nucleus \cite{Huang10,MaCW11PRC06IYR,MaCW12EPJA,MaCW12CPL06}. Based on the IYR method, a simple method is proposed to obtain the temperature using the $a_{sym}$ of the nucleus \cite{MA12CPL09asym} and $a_{sym}/T$ of the fragment (for hot primary IMFs \cite{Lin2014PRC,Liu2014PRC}, and for cold IMFs by considering the different $T$ dependence of the binding energy \cite{Ma2013NST}).

It is required that the reaction system should be in equilibrium in the thermodynamic models, which may be difficult to achieve in experiments or simulated reactions using dynamical models. The fragments may have different temperatures in the experiments. For example, in peripheral collisions the temperatures are different \cite{IBD13PRC,IBD15Tgt}, and it is also shown that temperature depends on the isospin of the reaction system \cite{MaCW12PRCT,MaCW13PRCT,SuJun11PRCTiso}. In this article, in the framework of the thermodynamics model, a thermometer is proposed to extract the temperature from the IMFs via the difference between IYRs, which is improved from the IYR method \cite{MaCW12PRCT,MaCW13PRCT}. The article is organized as follows. In Sec. \ref{ScTIB}, the improved isobaric ratio method for temperature is described. In Sec. \ref{RST}, the fragment yields in the measured $^{40, 48}$Ca and $^{58, 64}$Ni reactions, and the larger reaction systems of $^{124, 136}$Xe and $^{112, 124}$Sn are analyzed. The results of the temperatures from the IMFs are discussed. In Sec. \ref{summary} a summary of the article is presented.

\section{Isobaric ratio difference thermometer}
\label{ScTIB}
The new thermometer is developed using the canonical ensemble theory. Within the grand-canonical limitation, the cross section of a fragment $\sigma(A, I)$ has a form of \cite{GrandCan,Tsang07BET},
\begin{equation}\label{yieldGC}
\sigma(A, I) = CA^{\tau}\mbox{exp}\{[- F(A, I) + \mu_{n}N + \mu_{p}Z]/T\},
\end{equation}
where $C$ is a constant, $T$ is temperature, $\mu_n$ ($\mu_p$) is the chemical potential of neutrons (protons), $I \equiv N - Z$ is the neutron-excess, and $F(A, I)$ denotes the free energy, which depends on $T$ and can be parametrized as the $T$ dependent mass formula \cite{Huang10,LeeBE_T10PRC,MaCW11PRC06IYR,MaCW12EPJA,MaCW12CPL06}.

The isobaric yield ratio will be defined. For isobars with $I + 2$ and $I$, one has,
\begin{eqnarray}\label{RI2}
&\mbox{ln}R(A, I + 2, I)
=\mbox{ln}[\sigma(A, I + 2)/\sigma(A, I)]  \nonumber\\
& =[F(A, I) - F(A, I + 2) + \Delta\mu]/T,
\end{eqnarray}
with $\Delta\mu \equiv \mu_n-\mu_p$. Similarly, for isobars with $I$ and $I - 2$, one has,
\begin{eqnarray}\label{RIm2}
&\mbox{ln}R(A, I, I - 2)
=\mbox{ln}[\sigma(A, I)/\sigma(A, I - 2)]  \nonumber\\
& =[F(A, I - 2) - F(A, I) + \Delta\mu]/T.
\end{eqnarray}
Using different approximations of $F$, Eq. (\ref{RI2}) has been used to determine $T$ \cite{MaCW12PRCT,MaCW13PRCT}. In the IBD analysis, $\Delta\mu/T$ from the small $A$ fragment changes very little \cite{IBD13PRC,IBD13JPG,IBD14Ca,IBDCa48EPJA,IBD15Tgt,MaPLB15,Ma13finite,IBD15AMD}, which means that $\Delta\mu/T$ can be canceled out in the difference between the IYRs in Eqs. (\ref{RI2}) and (\ref{RIm2}). Thus, one obtains,
\begin{eqnarray}\label{DifRI}
&\mbox{ln}R(A, I + 2, I) - \mbox{ln}R(A, I, I - 2) \hspace{2.5cm} \nonumber\\
& =[2F(A, I) - F(A, I + 2) - F(A, I - 2)]/T,
\end{eqnarray}
Defining the residue free energy among the isobars as $\Delta F\equiv 2F(A, I) - F(A, I + 2) - F(A, I - 2)$, $T$ can be obtained once $\Delta F$ is known. For fragments having finite temperatures, it is proven that the residue free energy between two isobars can be replaced by that of $B(A, I)$ \cite{MaCW12PRCT}. A further secondary decay modification is also considered in a recent work \cite{MaCW13PRCT}. Following the assumption in Refs. \cite{MaCW12PRCT,MaCW13PRCT}, $B(A, I)$ will be used to replace $F(A, I)$.  From Eq. (\ref{DifRI}), the improved method to obtain $T$ from the difference between IYRs (labelled as $T_{IB}$) can be written as,
\begin{equation}\label{ImTIYRB}
T_{IB} = \frac{2B(A, I) - B(A, I + 2) - B(A, I - 2)}{\mbox{ln}R(A, I + 2, I) - \mbox{ln}R(A, I, I - 2)}.
\end{equation}
It is thus assumed that the residual free energy is equal to the residual binding energy for the three related isobars, i.e., $\Delta B \equiv 2B(A, I) - B(A, I + 2) - B(A, I - 2)$. $\Delta\mbox{ln}R \equiv \mbox{ln}R(A, I + 2, I) - \mbox{ln}R(A, I, I - 2)$ is defined to simplify the discussion of $T_{IB}$. Eq. (\ref{ImTIYRB}) is very similar to the isotopic temperature neglecting the spin term \cite{AlbNCA85DRT,Ma13CTP}. The binding energies in AME12 \cite{AME12} will be adopted in the analysis.

For the secondary decay modification of $T_{IB}$, an additional term considering the light particle decay is added to the binding energy \cite{Tsang07BET,MaCW13PRCT},
\begin{equation}\label{ModBnd}
F^{'} \approx -0.5\mbox{min}(S_n, S_p, S_{2n}, S_{2p}, S_{\alpha}) - E_{p},
\end{equation}
with $S_{i}$ denoting the separation energy of the corresponding particle, and $E_{p}$ being the modification of the pairing energy. In this article, only the odd $A$ fragment is considered, for which the pairing energy can be omitted. Though it has been suggested that for the odd $A$ fragments, no secondary decay modification is needed when using the binding energy \cite{MaCW13PRCT}, we will test this assumption in this work.

\begin{figure}[htbp]
\centering
\includegraphics
[width=8.6cm]
{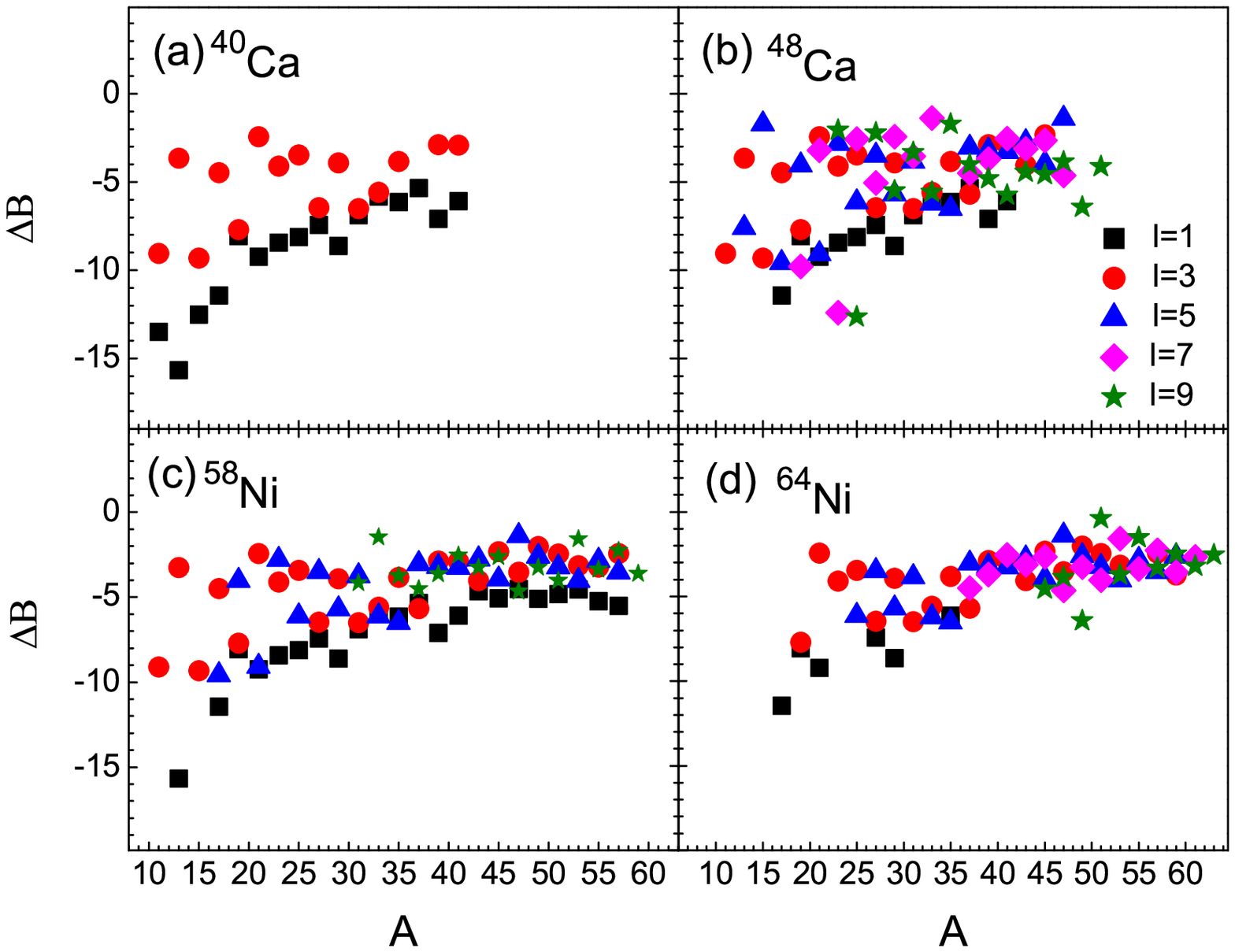}
\caption{\label{DeltaBfrag} (Color online) $\Delta B$ (in MeV) for the related isobars in the 140$A$ MeV $^{40, 48}$Ca + $^{9}$Be and $^{58, 64}$Ni + $^{9}$Be reactions. The binding energies for the ground state nuclei are adopted from Ref. \cite{AME12}.}
\includegraphics[width=8.6cm]
{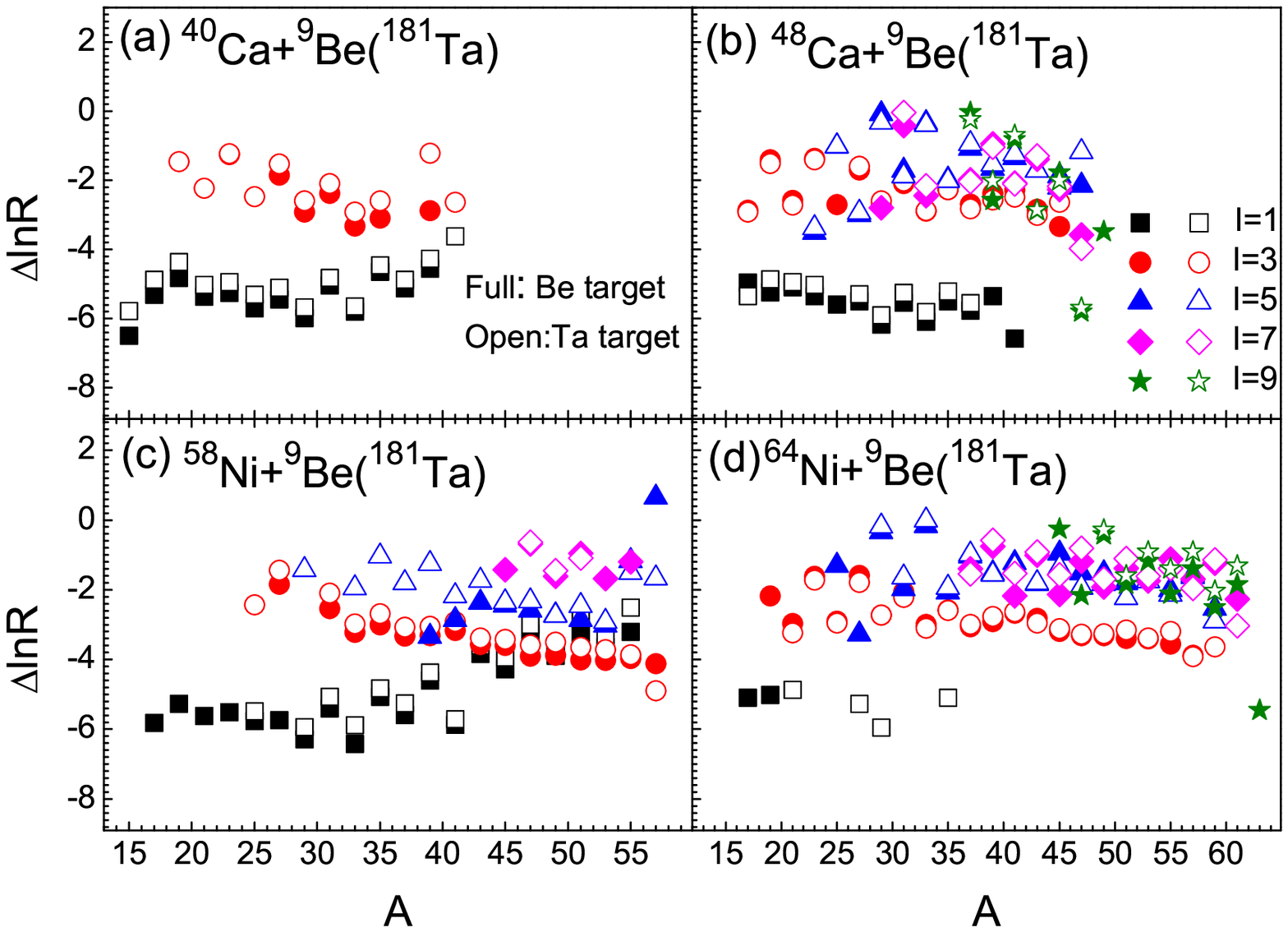}
\caption{\label{DeltaYfrag} (Color online) $\Delta \mbox{ln}R$ for the related isobars in the 140$A$ MeV $^{40, 48}$Ca ($^{58, 64}$Ni) + $^{9}$Be ($^{181}$Ta). The full and open symbols denote the results for the reactions with the $^{9}$Be and $^{181}$Ta targets, respectively.}
\end{figure}

\section{Results and discussion}
\label{RST}
The measured fragment yields in the reactions of 140$A$ MeV $^{40, 48}$Ca + $^{9}$Be ($^{181}$Ta) and $^{58, 64}$Ni + $^9$Be ($^{181}$Ta) \cite{Mocko06}, 1$A$ GeV $^{124, 136}$Xe + Pb \cite{Henz08} and $^{112,124}$Sn + $^{112,124}$Sn \cite{SnSndata}, will be adopted in the analysis.

\subsection{$\Delta B$ and $\Delta\mbox{ln}R$ distributions}

The 140$A$ MeV $^{40, 48}$Ca and $^{58, 64}$Ni projectile fragmentation reactions have been experimentally studied by Mocko {\it et al.} at the National Superconducting Cyclotron Laboratory at Michigan State University (MSU) \cite{Mocko06}. First, we studied the distributions of $\Delta B$ for fragments in the 140$A$ MeV $^{40, 48}$Ca + $^{9}$Be and $^{58, 64}$Ni + $^9$Be reactions (see Fig. \ref{DeltaBfrag}). The results are plotted separately in order to see the trend of $\Delta B$ more clearly. Though $\Delta B$ for the $I =$ 1 fragments almost changes monotonically with $A$, staggering is shown in the $\Delta B$ for $I =$ 3, 5, and 7 fragments on the relative small $A$ side. The staggering in $\Delta B$ becomes smaller for the $A >$ 35 fragments.

\begin{figure}[htbp]
\centering
\includegraphics
[width=8.6cm]
{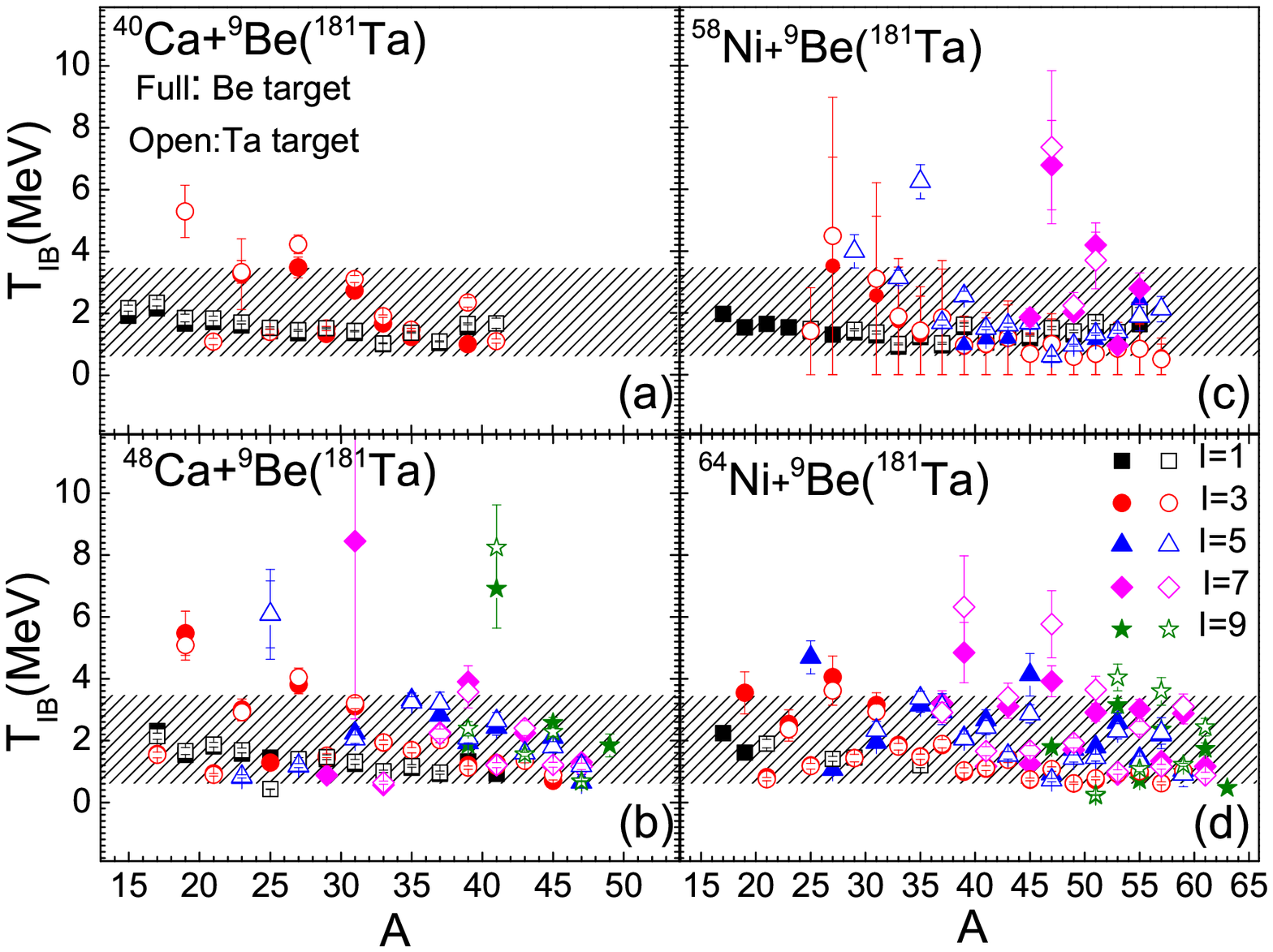}
\caption{\label{PimpT-CaNi} (Color online) $T_{IB}$ from isobaric yield ratio difference in the measured 140$A$ MeV $^{40}$Ca + $^{9}$Be ($^{181}$Ta) [in (a)], $^{48}$Ca + $^{9}$Be ($^{181}$Ta) [in (b)], $^{58}$Ni + $^{9}$Be ($^{181}$Ta) [in (c)], and $^{64}$Ni + $^{9}$Be ($^{181}$Ta) [in (d)] reactions. The results with the $^{9}$Be and $^{181}$Ta targets are denoted by the full and open symbols, respectively. The shadowed area denotes the range of 0.6 MeV $< T_{IB} < $ 3.5 MeV.
}
\includegraphics
[width=8.6cm]
{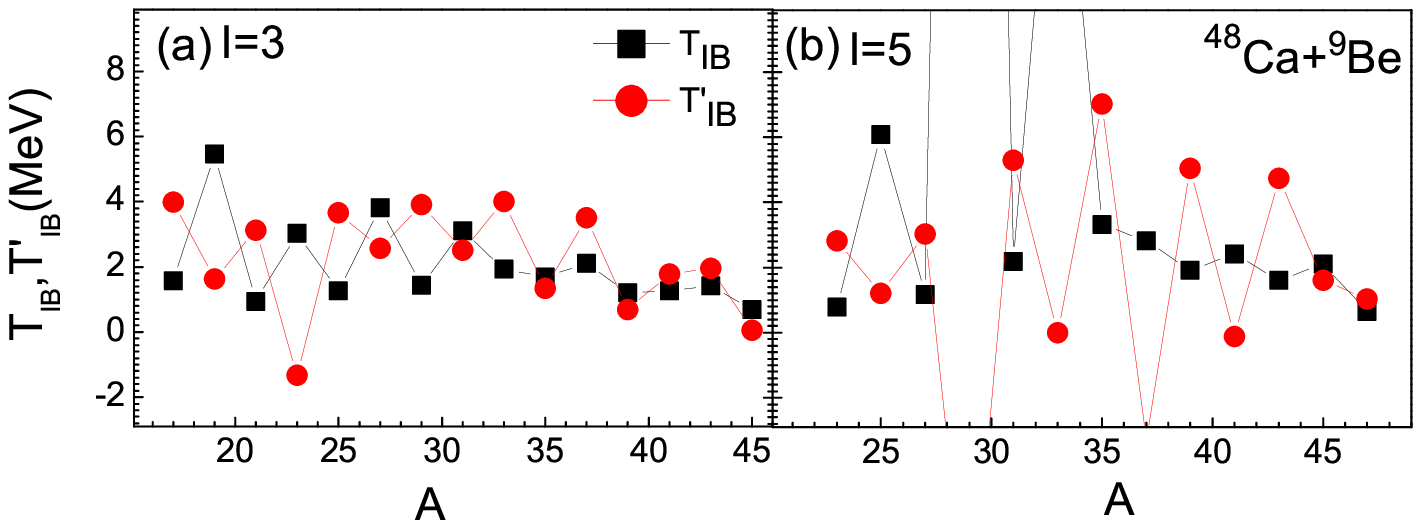}
\caption{\label{ModTIB} (Color online) A comparison between $T_{IB}$ and $T^{'}_{IB}$ of the $I = $ 3 and 5 fragments in the $^{48}$Ca + $^{9}$Be reactions. $T^{'}_{IB}$ is calculated by adding the secondary decay modification [Eq. (\ref{ModBnd})] to the binding energy in Eq. (\ref{ImTIYRB}).
}
\end{figure}

The $\Delta\mbox{ln}R$ for the related isobars in the 140$A$ MeV $^{40, 48}$Ca + $^{9}$Be($^{181}$Ta) and $^{58, 64}$Ni + $^9$Be($^{181}$Ta) reactions are plotted in Fig. \ref{DeltaYfrag}. For the $I = $ 1 fragments, $\Delta\mbox{ln}R$ is almost constant on the small $A$ side, but it increases with $A$ when $A > $ 40 and some staggering is shown when $A > $ 30. For the $I = $ 3, 5, and 7 fragments, an obvious staggering appears in $\Delta\mbox{ln}R$ on the small $A$ side, but this staggering becomes very small when $A$ is relative large. The target (Be and Ta) shows very little influence on the results of $\Delta\mbox{ln}R$. In general, the distributions of $\Delta B$ and $\Delta\mbox{ln}R$ are very similar in shape.

\subsection{$^{40, 48}$Ca ($^{58, 64}$Ni) + $^9$Be($^{181}$Ta) reactions}

$T_{IB}$ obtained from the fragments produced in the 140$A$ MeV $^{40, 48}$Ca + $^{9}$Be ($^{181}$Ta) and $^{58, 64}$Ni + $^{9}$Be ($^{181}$Ta) reactions has been plotted in Fig. \ref{PimpT-CaNi}. $T_{IB}$  from the $I = 1$ fragments is almost constant around 1.5 MeV in all the reactions. $T_{IB}$ from the $I = $ 3 fragments shows a relatively large staggering for the small $A$ fragments, but it becomes similar to that from the $I = $1 fragments when $A > \sim$ 35. $T_{IB}$ from the $I =$ 5 fragments shows a small staggering. But the staggering phenomenon again appears in $T_{IB}$ for $I =$ 7 and 9 fragments. It is also shown that $T_{IB}$ slightly depends on $I$. From most of the fragments with $I > $3, the values of $T_{IB}$ are within a range from 0.6 MeV to 3.5 MeV (the shadowed areas), which agrees with the temperatures obtained by the IYR method \cite{MaCW12PRCT,MaCW13PRCT}. Only for some of the very-neutron rich fragments, the values of $T_{IB}$ are large. In the canonical ensemble theory, $T =$ 2.2 MeV has been used to estimate the mass of neutron-rich copper isotopes \cite{Tsang07BET}. This agrees with the results in this work. As to the target effect, a relatively large difference appears in the $T_{IB}$ of large-$I$ fragments ($I =$ 7 and 9) for the reactions using the $^{9}$Be and $^{181}$Ta targets.

\begin{figure}[htbp]
\centering
\includegraphics
[width=8.6cm]
{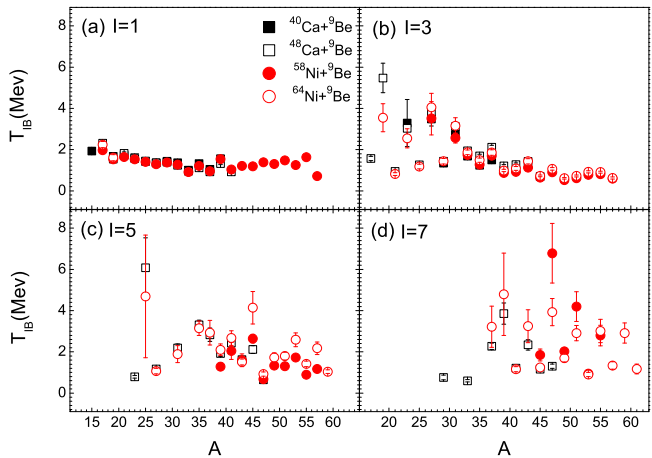}
\caption{\label{TIBCaNiI} (Color online) A comparison of $T_{IB}$ from fragments in the 140$A$ MeV $^{40, 48}$Ca + $^{9}$Be and $^{58 ,64}$Ni + $^{9}$Be reactions according to the neutron-excess $I$, which are re-plotted from the results in Fig. \ref{PimpT-CaNi}.
}
\end{figure}

Now we study the secondary decay modification of $T_{IB}$. The isotopic thermometer does not consider the secondary decay modification \cite{Ma13CTP}. In fact, for light particles, there is no need to consider the secondary decay modification. While for the IMFs, it should be verified whether the secondary decay modification is needed. At the time the fragment is formed, the system is assumed to be in equilibrium with a uniform temperature. The secondary decay modification is introduced with the aim of making the $T_{IB}$ more consistent, rather than enlarge the fluctuation. In this work, by adding the secondary decay modification [Eq. (\ref{ModBnd})] term to Eq. (\ref{ImTIYRB}), a modified temperature $T^{'}_{IB}$ can be obtained.  $T^{'}_{IB}$ from the $I = $3 and 5 fragments in the $^{48}$Ca + $^9$Be is plotted in Fig. \ref{ModTIB}. It is seen that the secondary decay modification does not lead to a more consistent temperature, and larger fluctuation appears in $T^{'}_{IB}$. The results suggest that no secondary decay modification is needed for the odd $I$ (or $A$) fragments, which agrees with the assumption in the isobaric ratio thermometer \cite{MaCW13PRCT}.

\begin{figure}[htbp]
\centering
\includegraphics[width=8.6cm]
{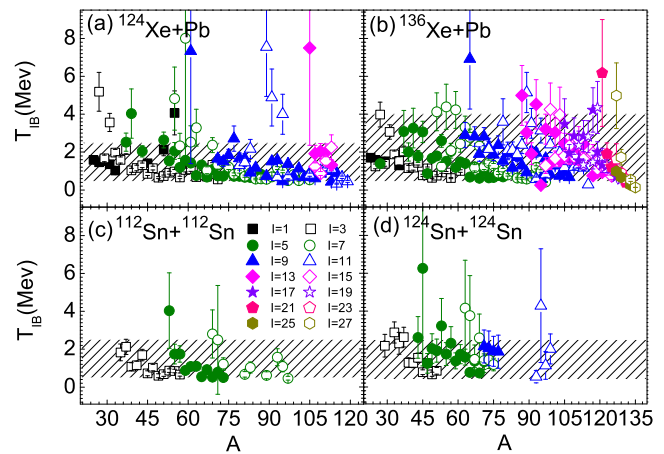}
\caption{\label{TIBXeSn} (Color online) $T_{IB}$ of fragments in the measured 1$A$ GeV $^{124}$Xe + Pb [in (a)], $^{136}$Xe + Pb [in (b)] \cite{Henz08}, $^{112}$Sn + $^{112}$Sn [in (c)], and $^{124}$Sn + $^{124}$Sn [in (d)] reactions \cite{SnSndata} at FRS of GSI. The shadowed area denotes the range of 0.5 MeV $< T_{IB}< $ 2.5 MeV in panels (a), (c) and (d), but 0.5 MeV $< T_{IB}< $ 4 MeV in panel (b).
}
\end{figure}

In the isoscaling or the IBD methods, the fragments in the two reactions are assumed to have the same temperature, which can make it possible to cancel out the free energy of the fragments \cite{IBD13PRC,IBD14Ca,IBD15AMD} (in the recent Shannon information uncertainty method, this is not required \cite{MaPLB15}). Since $T_{IB}$ is obtained from fragments, it is interesting to check whether $T_{IB}$ is the same in the different reactions. The values of $T_{IB}$ from the fragments in the $^{40, 48}$Ca + $^9$Be and $^{58, 64}$Ca + $^9$Be reactions are re-plotted according to $I$ in Fig. \ref{TIBCaNiI} for a comparison. It can be seen that in the four reactions, $T_{IB}$ from the $I =$ 1 fragments is almost the same [Fig. \ref{TIBCaNiI}(a)], and it is also almost the same for the $I =$ 3 fragments, except those having a relatively small $A$ [Fig. \ref{TIBCaNiI}(b)]. While an obvious difference appears in $T_{IB}$ for the $I = $ 5 fragments, and the difference becomes even larger in the $I = $ 7 fragments. It is also shown that $T_{IB}$ for the $I =$ 5 fragments is quite similar in the neutron-rich $^{48}$Ca and $^{64}$Ni reactions. These results suggest that for the isoscaling and IBD methods, the temperature can be assumed as the same for the fragment which does not have a very large neutron-excess.

\begin{figure*}[htbp]
\centering
\includegraphics
[width=15cm]
{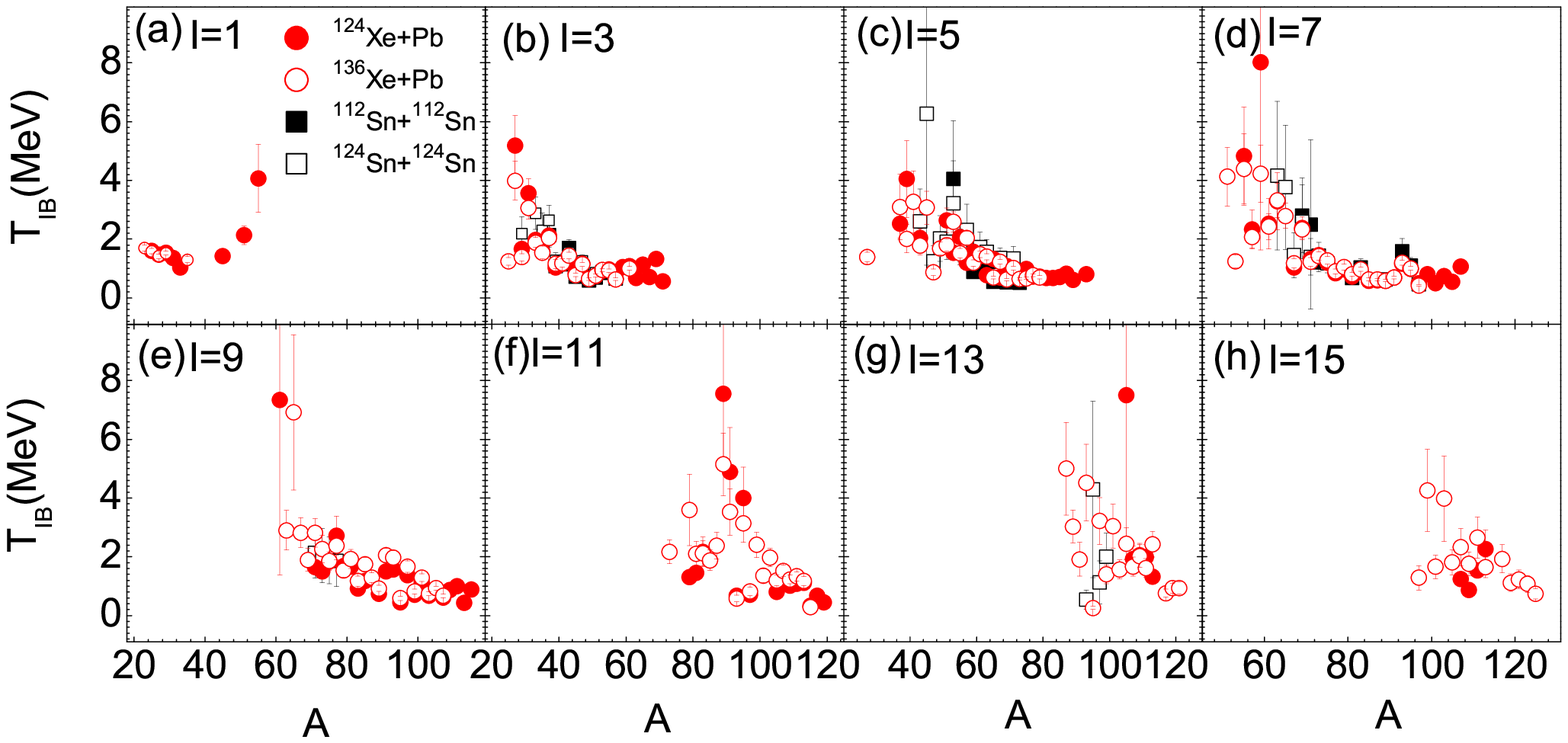}
\caption{\label{TIBIcompXeSn} (Color online) Comparison of $T_{IB}$ from fragments with the same $I$ in the measured 1$A$ GeV $^{124, 136}$Xe + Pb and $^{112, 124}$Sn + $^{112, 124}$Sn reactions.
}
\includegraphics
[width=15cm]
{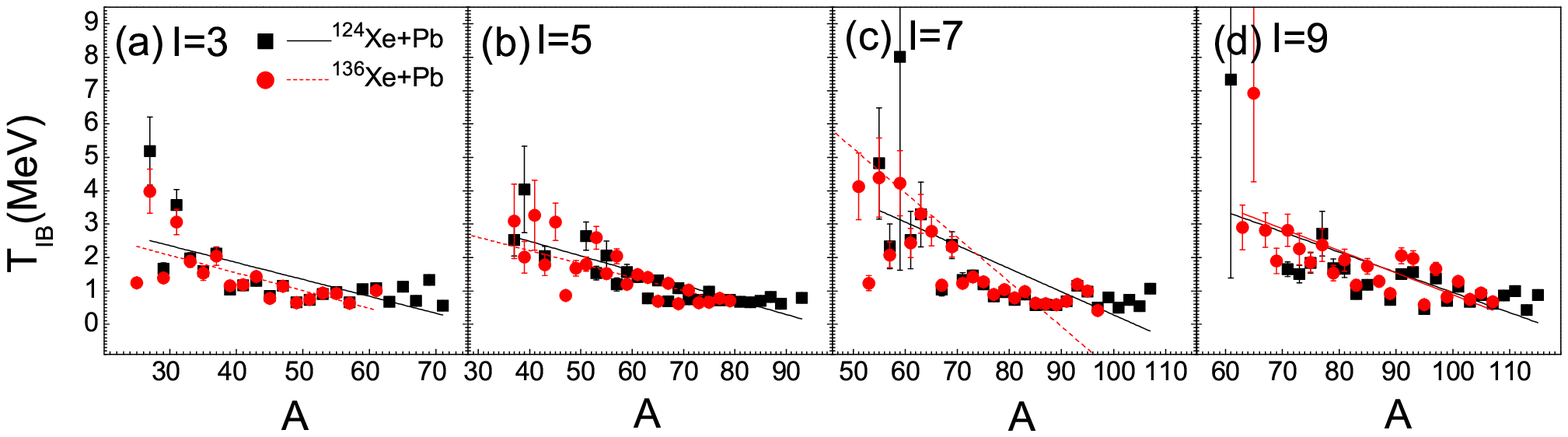}
\caption{\label{Tfit} (Color online) Fitting results (lines) for the correlation between $T_{IB}$ and $A$ of the $I =$ 3, 5, 7, and 9 fragments in the $^{124, 136}$Xe reactions using a function $y = C (1- kA)$. The fitted $C$ and $k$ are plotted in Fig. \ref{fitresult}.
}
\end{figure*}

\subsection{1$A$ GeV $^{124, 136}$Xe + Pb reactions}

The fragment yields in the 1$A$ GeV $^{124, 136}$Xe + Pb projectile fragmentation reactions have been measured by Henzlova \textit{et al.} at the Fragment Separator (FRS) at GSI \cite{Henz08}. The measured fragments cover a broad range of isotopes of the elements between $Z =$ 3 and $Z = 56$ for $^{136}$Xe and between $Z =$ 5 and $Z = 55$ for $^{124}$Xe. The temperature from the the IYR method has been obtained \cite{MaCW12PRCT,MaCW12PRCT}, and the isoscaling and IBD results have also been reported \cite{IBD13JPG}. Since the Xe reaction systems are much larger than the Ca and Ni reactions, the IBD results are sensitive to the shell closure of the fragments \cite{IBD13JPG}.

The values of $T_{IB}$ from fragments in the 1$A$ GeV $^{124}$Xe + Pb and 1$A$ GeV $^{136}$Xe + Pb reactions have been plotted in Fig. \ref{TIBXeSn} (a) and (b), respectively. From most of the fragments in the $^{124}$Xe reactions, $T_{IB}$ falls in a narrower range of 0.5 MeV $< T_{IB} <$ 2.5 MeV, compared to that of 0.5 MeV $< T_{IB} <$ 4 MeV in the $^{136}$Xe reactions, as shown by the shadowed areas. Some staggering is shown in $T_{IB}$ from the small $A$ fragments where $I = $ 3, 5, and 7 in both of the reactions, and in the fragments where $I = $ 13 and 15 in the $^{136}$Xe reaction. From the fragments with the same $I$, $T_{IB}$ decreases apparently with the increasing $A$, while $T_{IB}$ from the relatively large $A$ fragment tends to be constant.

\subsection{1$A$ GeV $^{112, 124}$Sn + $^{112, 124}$Sn reactions}

The fragments with atomic numbers $Z > $ 10 in the 1$A$ GeV $^{112,124}$Sn + $^{112,124}$Sn reactions have also been measured at FRS at GSI by F\"{o}hr \textit{et al.} \cite{SnSndata}. The cross sections of the fragments in these reactions have been studied in theories, such as the statistical multifragmentation model (SMM) \cite{Sn-SMM15}, and the EPAX3 parametrizations \cite{EPAX3}. The temperatures of the fragments have been analyzed by using the isobaric ratio methods \cite{MaCW12PRCT,MaCW12PRCT}. The results of the isoscaling and IBD methods also have been reported \cite{IBD13JPG}.

The values of $T_{IB}$ from the fragments in the 1$A$ GeV $^{112}$Sn + $^{112}$Sn and $^{124}$Sn + $^{124}$Sn reactions have been plotted in Fig. \ref{TIBXeSn} (a) and (b), respectively. $T_{IB}$ from the fragments in the $^{112, 124}$Sn reactions have the similar trend to those in the $^{124, 136}$Xe reactions, most of which are in the range 0.5 MeV $< T_{IB} <$ 2.5 MeV.

The values of $T_{IB}$ from the fragments in the four reactions are compared according to $I$, as shown in Fig. \ref{TIBIcompXeSn}. For the fragments with the same $I$, $T_{IB}$ is very similar, which indicates that in the isoscaling and IBD methods, it is reasonable to assume that the temperatures of a specific fragment in two reactions are the same.

\subsection{Discussion}
\label{disc}

The advantage of the $T_{IB}$ thermometer is that the temperature can be directly obtained from the IYR difference. Without knowledge of free energy, the residual binding energy for the isobars is used, which makes the analysis much easier. In addition, the fitting procedure in the IYR method \cite{MaCW12PRCT,MaCW13PRCT} is avoided, which makes $T_{IB}$ a direct probe to the temperature. Besides, the $T_{IB}$ thermometer also avoids the complexity of the coefficient ratio method from the IMFs \cite{Ma2013NST,Liu2014PRC}. The results of $T_{IB}$ can help to separate the temperature term in the isoscaling, IYR, and IBD methods.

\begin{figure}[htbp]
\centering
\includegraphics
[width=8.6cm]
{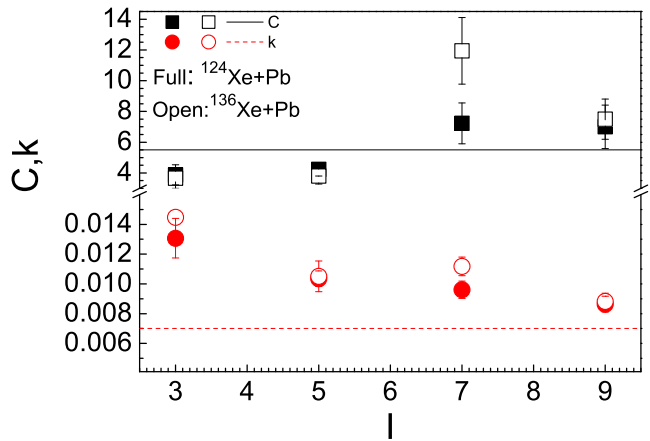}
\caption{\label{fitresult} (Color online) The values of $C$ (squares) and $k$ (dots) in the fitting function  $y = C (1- kA)$. The fitting is for the correlations between $T_{IB}$ and $A$ of the $I =$ 3, 5, 7, and 9 fragments in the $^{124, 136}$Xe reactions. The full and open symbols are for the $^{124}$Xe and $^{136}$Xe reactions, respectively. The solid and dashed lines denote the $C = $ 5.5 and $k =$ 0.007 for the primary fragments formed in the emitting source as in Ref. \cite{Liu2014PRC}.
}
\end{figure}

For hot prefragments, a mass dependence of the temperature is illustrated, i.e., $T = C (1 - kA)$ with $C= 5.5 \pm 0.2$ and $k=$ 0.007, with $C$ and $k$ variables with the neutron-excess of a fragment \cite{Liu2014PRC}. Due to the staggering phenomenon in $T_{IB}$, no obvious mass dependence of $T_{IB}$ is shown in the $^{40, 48}$Ca and $^{58, 64}$Ni reactions, but an obvious mass dependence of $T_{IB}$ is shown in the $^{124, 136}$Xe and $^{124, 136}$Sn reactions. The $T_{IB}$ from the $I =$ 3, 5, 7, and 9 fragments in the $^{124, 136}$Xe reactions have been fitted using the function $y = C (1- k\cdot A)$ (see Fig. \ref{Tfit}). The fitting function in general coincides with the data, except for the staggering. $C$ and $k$ (see Fig. \ref{fitresult}) both depend on $I$, with $C$ slightly increasing with $A$ and $k$ slightly decreasing as $A$ increases. Meanwhile, for the two reactions, the values of $C$ and $k$ are similar. The values of $C$ fitted in this work are close to those of the hot fragments, while the values of $k$ in this work are larger than those of the hot fragments \cite{Liu2014PRC}, which means that $T_{IB}$ from the cold fragments depends more on $A$ than those from the primary ones.

\section{Summary}
\label{summary}

To summarize, an improved isobaric ratio thermometer ($T_{IB}$) for IMFs has been developed based on the difference between IYRs, in which the residual binding energy is used instead of the residue free energy. In contrast to the IYR thermometer, $T_{IB}$ is obtained directly from fragments and avoids the fitting procedure in the IYR method, which makes $T_{IB}$ become a direct probe to temperature. Considering no secondary decay modification, $T_{IB}$ from the odd $I$ fragments in the 140$A$ MeV $^{40, 48}$Ca + $^{9}$Be ($^{181}$Ta) and $^{58, 64}$Ni + $^{9}$Be ($^{181}$Ta), and the 1$A$ GeV $^{124, 136}$Xe + Pb and $^{112,124}$Sn + $^{112,124}$Sn reactions, has been obtained. The values of $T_{IB}$ for most considered IMFs are low. It is also concluded that, for similar reactions with different asymmetries, $T_{IB}$ can be assumed as the same, which satisfies the assumption that the temperature in two similar reactions should be the same in the isoscaling, IYR, and IBD methods. $T_{IB}$ also shows a slight dependence on the $A$ of a fragment, which is reflected in the temperature from the primary fragment.

\begin{acknowledgments}
This work is supported by the Program for Science and Technology Innovation Talents in Universities of Henan Province (13HASTIT046). X.-G. Cao thanks the  National Natural Science Foundation of China (grant no. 11305239) for support.
\end{acknowledgments}


\begin{thebibliography}{}

\bibitem{YGMaZipfPRL99}
Y. G. Ma, Phys. Rev. Lett. \textbf{83}, 3617 (1999).
\bibitem{YgMa99PRCTc}
Y. G. Ma 
{\it et al.},
Phys. Rev. C \textbf{60}, 024607 (1999).
\bibitem{EOSPRC02Tc}
B. K. Srivastava \textit{et al.} (EOS Collaboration),
Phys. Rev. C \textbf{65}, 054617 (2002).

\bibitem{DasPRC02Tc}
C. B. Das 
{\it et al.},
Phys. Rev. C \textbf{66}, 044602 (2002).

\bibitem{YgMa04PRCTc}
Y. G. Ma 
{\it et al.},
Phys. Rev. C \textbf{69}, 031604(R) (2004).

\bibitem{YgMa05PRCTc}
Y. G. Ma 
{\it et al.},
Phys. Rev. C \textbf{71}, 054606 (2005).
\bibitem{AlbNCA85DRT}
S. Albergo 
{\it et al.},
Nuovo Cimento A {\bf 89}, 1 (1985).

\bibitem{Nato95T}
J. B. Natowitz 
{\it et al.},
Phys. Rev. C \textbf{52}, R2322 (1995).

\bibitem{Poch95PRLDR_T}
J. Pochodzalla 
\textit{et al.},
Phys. Rev. Lett. \textbf{75}, 1040 (1995).

\bibitem{JSWangT05}
J. Wang 
{\it et al.},
Phys. Rev. C \textbf{72}, 024603 (2005).

\bibitem{Wada97T}
R. Wada 
{\it et al.},
Phys. Rev. C \textbf{55}, 227 (1997).

\bibitem{Surfling98TPRL}
V. Serfling 
{\it et al.},
Phys. Rev. Lett. \textbf{80}, 3928 (1998).

\bibitem{XiHF98Tiso}
H. F. Xi 
\textit{et al.},
Phys. Rev. C \textbf{58}, 2636 (1998).
\bibitem{Odeh2000PRLT}
T. Odeh 
\textit{et al.,}
Phys. Rev. Lett. \textbf{84}, 4557 (2000).

\bibitem{TrautTHFrg07}
W. Trautmann \textit{et al.} (ALADIN Collaboration),
Phys. Rev. C {\bf 76}, 064606 (2007).
\bibitem{Ma13CTP} 
C. W. Ma \textit{et al.},
Commun. Theor. Phys. \textbf{59}, 95 (2013).
\bibitem{Bonasera11PLBT}
H. Zheng, A. Bonasera, Phys. Lett. B \textbf{696}, 178 (2011).

\bibitem{ZhouPei11T}
P. Zhou 
{\it et al.},
Phys. Rev. C \textbf{84}, 037605 (2011).

\bibitem{Morr84PLBT_E}
D. J. Morrissey %
\textit{et al.},
Phys. Lett. B \textbf{148}, 423 (1984).

\bibitem{Poch85PRL_E}
J. Pochodzalla %
\textit{et al.},
Phys. Rev. Lett. \textbf{55}, 177 (1985).

\bibitem{Nato02PRCT_E}
J. B. Natowitz 
\textit{et al.},
Phys. Rev. C \textbf{65}, 034618 (2002).
\bibitem{flucT10} 
S. Wuenschel 
\textit{et al.},
Nucl. Phys. A \textbf{843}, 1 (2010).
\bibitem{Westfall82PLBT_spec}
G. D. Westfall, Phys. Lett. B \textbf{116}, 118 (1982).

\bibitem{Jacak83T_spec}
B. V. Jacak %
\textit{et al.},
Phys. Rev. Lett. \textbf{51}, 1846 (1983).

\bibitem{SuJPRC12T_spec}
J. Su 
{\it et al.},
Phys. Rev. C \textbf{85}, 017604 (2012).
\bibitem{Huang10}
M. Huang 
{\it et al.},
Phys. Rev. C \textbf{81}, 044620 (2010).

\bibitem{Huang10PRCTf}
M. Huang 
{\it et al.},
Phys. Rev. C \textbf{82}, 054602 (2010).
\bibitem{ModelFisher1} 
R. W. Minich, S. Agarwal, A. Bujak 
{\it et al.},
Phys. Lett. B \textbf{118}, 458 (1982).
\bibitem{ModelFisher3}
A. S. Hirsch 
{\it et al.},
Phys. Rev. C \textbf{29}, 508 (1984).

\bibitem{Tsang07BET} 
M. B. Tsang 
\textit{et al.},
Phys. Rev. C \textbf{76}, 041302(R) (2007).
\bibitem{LeeBE_T10PRC}
S. J. Lee and A. Z. Mekjian,
Phys. Rev. C \textbf{82}, 064319 (2010).
\bibitem{Lin2014PRC}
W. Lin, X. Liu, M. R. D. Rodrigues
\textit{et al.},
Phys. Rev. C \textbf{89}, 021601(R) (2014); \textit{ibid}, \textbf{90}, 044603 (2014).
\bibitem{Liu2014PRC} 
X. Liu 
\textit{et al.},
Phys. Rev. C \textbf{90}, 014605 (2014); \textit{ibid}, \textbf{92}, 014623 (2015);
\textit{ibid.}, Nucl. Phys. A \textbf{933}, 290 (2015);
X. Liu, M. Huang, R. Wada, Nucl. Sci. Tech. \textbf{26}, S20508 (2015). 

\bibitem{MaCW12PRCT} 
C. W. Ma, J. Pu, Y. G. Ma, R. Wada, S. S. Wang,
Phys. Rev. C \textbf{86}, 054611 (2012).
\bibitem{MaCW13PRCT}
C. W. Ma, X. L. Zhao, J. Pu 
{\it et al.},
Phys. Rev. C \textbf{88}, 014609 (2013).

\bibitem{Ma2013NST}  
C. W. Ma, C. Y. Qiao, S. S. Wang 
\textit{et al.},
Nucl. Sci. Tech. \textbf{24}, 050510 (2013).



\bibitem{MBTsPRL01iso}
M. B. Tsang 
\textit{et al.},
Phys. Rev. Lett. \textbf{86}, 5023 (2001).

\bibitem{Botv02PRCiso}
A. S. Botvina 
\textit{et al.},
Phys. Rev. C \textbf{65}, 044610 (2002).

\bibitem{Ono03PRCiso}
A. Ono, P. Danielewicz, W. A. Friedman, W. G. Lynch, and M. B. Tsang,
Phys. Rev. C \textbf{68}, 051601(R) (2003).

\bibitem{Ono04RRCiso}
A. Ono, P. Danielewicz, W. A. Friedman, W. G. Lynch, and M. B. Tsang,
Phys. Rev. C \textbf{70}, 041604(R) (2004).

\bibitem{SouzaPRC09isot} 
S. R. Souza, M. B. Tsang, B. V. Carlson, R. Donangelo, W. G. Lynch, and A. W. Steiner,
Phys. Rev. C \textbf{80}, 044606 (2009).

\bibitem{MaCW11PRC06IYR}
C. W. Ma, F. Wang, Y. G. Ma, and C. Jin,
Phys. Rev. C \textbf{83}, 064620 (2011).

\bibitem{MaCW12EPJA}
C.-W. Ma 
{\it et al.},
Eur. Phys. J. A \textbf{48}, 78 (2012). 

\bibitem{MaCW12CPL06}
C.-W. Ma 
{\it et al.},
Chin. Phys. Lett. \textbf{29}, 062101 (2012).

\bibitem{Ma13finite} 
C. W. Ma, S. S. Wang, H. L. Wei, and Y. G. Ma,
Chin. Phys. Lett. \textbf{30}, 052101 (2013);
C. W. Ma, H. L. Wei, Y. G. Ma, Phys. Rev. C \textbf{88}, 044612 (2013).
\bibitem{RCIMa15}  
C.-W. Ma, S.-S. Wang, Y.-L. Zhang, H.-L. Wei, Commun. Theor. Phys. \textbf{64}, 334 (2015).

\bibitem{IBD13PRC}
C. W. Ma, S. S. Wang, Y. L. Zhang, H. L. Wei, Phys. Rev. C \textbf{87}, 034618 (2013).
\bibitem{IBD13JPG}
C. W. Ma, S. S. Wang, Y. L. Zhang, H. L. Wei, J. Phys. G: Nucl. Part. Phys. \textbf{40}, 125106 (2013).
\bibitem{IBD14Ca}
C. W. Ma, J. Yu, X. M. Bai, Y. L. Zhang, H. L. Wei, S. S. Wang,
Phys. Rev. C \textbf{89}, 057602 (2014).
\bibitem{IBDCa48EPJA}
C. W. Ma, X. M. Bai, J. Yu, H. L. Wei,
Eur. Phys. J. A \textbf{50}, 139 (2014).
\bibitem{YMNST15}
M. Yu \textit{et al.}, Nucl. Sci. Tech. \textbf{26}, S20503 (2015).
\bibitem{IBD15Tgt}
C.-W. Ma, Y.-L. Zhang, C.-Y. Qiao, S.-S. Wang,
Phys. Rev. C \textbf{91}, 014615 (2015).
\bibitem{IBD15AMD}
C.-Y. Qiao, H.-L. Wei, C.-W. Ma 
\textit{et al.},
Phys. Rev. C \textbf{92}, 014612 (2015).

\bibitem{MaPLB15} C. W. Ma 
\textit{et al.},
Phys. Lett. B \textbf{742}, 19 (2015).

\bibitem{MA12CPL09asym}
C. W. Ma 
{\it et al.},
Chin. Phys. Lett. \textbf{29}, 092101 (2012).

\bibitem{SuJun11PRCTiso}
J. Su and F. S. Zhang, Phys. Rev. C {\bf 84}, 037601 (2011).
\bibitem{GrandCan} 
C. B. Das, S. Das Gupta, X. D. Liu, and M. B. Tsang,
Phys. Rev. C \textbf{64}, 044608 (2001).

\bibitem{AME12}
M. Wang, G. Audi, A. H. Wapstra,
\textit{et al.},
Chin. Phys. C \textbf{36}, 1603 (2012).

\bibitem{Mocko06}
M. Mocko 
{\it et al.},
Phys. Rev. C \textbf{74}, 054612 (2006).
\bibitem{Henz08}
D. Henzlova \textit{et al.}, Phys. Rev. C \textbf{78}, 044616 (2008).
\bibitem{SnSndata}
V. F\"{o}hr \textit{et al.}, Phys. Rev. C \textbf{84}, 054605 (2011).

\bibitem{Sn-SMM15}
H. Imal, A. Ergun, N. Buyukcizmeci, R. Ogul, A. S. Botvina, and W. Trautmann,
Phys. Rev. C \textbf{91}, 034605 (2015).
\bibitem{EPAX3}
K. S\"{u}mmerer, Phys. Rev. C \textbf{86}, 014601 (2012).

\end{thebibliography}
\end{document}